\documentstyle[12pt,aaspp4,epsfig]{article}
\def\spose#1{\hbox to 0pt{#1\hss}}

\def\msun{{\rm ~M}_{\odot}}

\def\beq{\begin{equation}}
\def\enq{\end{equation}}
\def\lta{\mathrel{\spose{\lower 3pt\hbox{$\mathchar"218$}}
     \raise 2.0pt\hbox{$\mathchar"13C$}}}
\def\gta{\mathrel{\spose{\lower 3pt\hbox{$\mathchar"218$}}
     \raise 2.0pt\hbox{$\mathchar"13E$}}}

\def\be{\begin{equation}}
\def\ee{\end{equation}}

\begin{document}

\title{Pulsar Wind Nebulae and the Non-thermal X-Ray Emission
of Millisecond Pulsars}

\author{K. S. Cheng\altaffilmark{1}, Ronald E. Taam\altaffilmark{2},
and W. Wang\altaffilmark{3}}

\affil{$^{1}$ Department of Physics,  University of Hong Kong,
       Pokfulam Road, Hong Kong, China\\
$^{2}$ Northwestern University, Dept. of Physics \& Astronomy,
2145 Sheridan Rd., Evanston, IL 60208, USA\\
$^{3}$ Max-Planck-Institut f\"ur extraterrestrische Physik,
Postfach 1312, 85741 Garching, Germany }

\begin{abstract}
The non-thermal, non-pulsed X-ray emission of MSPs
is investigated. As in young pulsars, MSPs emit a 
relativistic wind, which interacting with the interstellar medium 
and/or a binary companion can significantly contribute to the 
non-pulsed emission of these pulsars.  An application and extension 
of a simple model developed for young pulsars is applied to the old 
recycled MSP B1957+20.  It is found that the pulsar 
wind can, indeed, contribute to both the resolved and unresolved 
X-ray emission.  For other MSP in the Galactic 
field where the spectral index of the non-pulsed component has
been measured (i.e., PSR B1937+21, J0218+4232) the 
contribution of the pulsar wind to the non-pulsed X-ray luminosity 
is estimated.  For the MSPs in the core regions 
of globular clusters, the pulsar wind nebula is likely affected by its 
interaction with the dense stellar environment, possibly leading to 
a diminished contribution to the total X-ray emission.  In this case, the existence of 
non-thermal non-pulsed X-ray emission is more likely for binary rather 
than isolated MSPs with the emission arising from 
the interaction of the relativistic pulsar wind with a binary companion. 
Our study suggests that the magnetization parameter in the pulsar wind
nebulae of MSPs is significantly larger than that of the Crab 
nebula by about a factor of 10. The 
emission from MSPs moving at high velocities may 
appear spatially extended with a tail-like morphology, which may 
contribute to the faint 
filamentary X-ray source subpopulation in the Galaxy.
\end{abstract}


\section{INTRODUCTION}
Millisecond pulsars have been a subject of active interest ever
since the discovery of PSR 1937+21 by Backer et al. (1982).  It is
generally accepted that the neutron stars in these systems have
been spun up by accretion torques in binary systems (e.g., Alpar et al. 
1984; Phinney \& Kulkarni 1994).  This class of rotation powered pulsars can
remain active for as long as a Hubble timescale, offering the
opportunity for investigating the emission processes of old
recycled neutron stars.  Of interest in recent years have been
studies of the X-ray properties of these objects. Specifically,
pioneering studies carried out by Becker \& Tr\"umper (1999)
using ROSAT led to the detection of 9 millisecond pulsars
in the energy band from 0.1 - 2.4 keV.  Since little
X-ray emission is expected from the cooling of the neutron star,
due to the millisecond pulsars' long age (e.g., Tsuruta 1998),
investigators have primarily focused on emission processes taking
place in the magnetosphere (e.g., Cheng \& Ruderman 1980; Arons
1981; Zhang \& Harding 2000; Harding \& Muslimov 2001, 2002; Zhang \& Cheng 2003).

However, it has long been recognized that the nebulae surrounding
young pulsars are important for understanding the nature and
spatial extent of the non-thermal X-ray emission of these compact
objects after X-ray detections of the Crab nebula
(Peterson \& Jacobson 1970 and references therein).
In particular, the recent X-ray images of young pulsars obtained
with the {\em Chandra} Observatory have revealed spatially
resolved structures, which provide important diagnostic
information about the geometry of the inner regions of the Crab 
supernova remnant (Weisskopf et al. 2000) and Vela nebula
(Helfand, Gotthelf, \& Halpern 2001). Such extended structures are
not limited to young pulsars as they have also been observed 
from the three million year old pulsar PSR 1929+10 
(Becker et al. 2005) and from
the old binary millisecond pulsar PSR 1957+20 (Stappers et al.
2003).  The information gleaned from these observational studies
is vital for facilitating an understanding of the mechanism
involved in the production of the X-ray emission. For the majority
of pulsars, where the nebulae are unresolved, the wind nebulae
would contribute to the non-thermal non-pulsed X-ray emission
component (see Kennel \& Coroniti 1984; Chevalier 2000). Recently,
Cheng, Taam, \& Wang (2004) have found that the X-ray
properties of the non-thermal non-pulsed emission component of all
known rotation powered X-ray pulsars could be interpreted within
such a framework whether or not a wind nebula has been identified
with the pulsar.

In this paper, we build on the seminal study of the X-ray emission
properties of millisecond pulsars elucidated in Becker \&
Tr\"umper (1999) to examine the properties of the non-thermal
non-pulsed component from recycled neutron stars.  The origin of
the X-ray emission from these objects may be on the neutron star
surface, in the magnetosphere, and in the wind nebula.  Since
different radiation processes are involved in these spatially
distinct regions, it is likely that their contribution to the
total luminosity and energy spectra differ.  Cheng et al. (2004)
argue that the main contribution to the non-thermal non-pulsed
component in pulsars is due to the emission from the wind nebula.
In this paper, we assume this to be the case as well, although 
one cannot exclude magnetospheric contributions at luminosity 
levels lower than $\sim 3 \times 10^{-4}$ {of the spin down power}.  

Our interest in the X-ray emission of millisecond pulsars has been
further stimulated by the recent detection of faint X-ray point
sources with luminosities in the range from $\sim 10^{31} -
10^{35}$ ergs s$^{-1}$ in deep observational surveys of the
Galactic center by Wang, Gotthelf, \& Lang (2002) and Muno et al.
(2003). The spectra of the Galactic center sources in the innermost 
20 pc have been fitted to a power law of photon index, $\Gamma$, 
ranging from about -1 to 3 (Muno et al.  2003).  Muno et al. (2004) 
suggest that magnetic cataclysmic variables can
substantially contribute to the faint population characterized by
hard spectra.  On the other hand, these binary systems do not
significantly contribute to the fainter X-ray population
characterized by softer spectra. Belczynski
\& Taam (2004) suggest that among neutron star models, the neutron
stars in the quiescent state of soft X-ray transients in Roche
lobe overflow systems can contribute to the faint ($10^{31} -
10^{33}$ ergs s$^{-1}$), soft sources detected in the Muno et al.
(2003) survey.  The millisecond pulsars, which are likely 
descendants of such systems, may contribute more significantly to the general
population of such sources in the Galaxy because of their long
lifetimes.  Estimates of the number of millisecond pulsars in the
Galaxy exceeds $3 \times 10^4$ (Lyne et al. 1998).

Here, we explore millisecond pulsars as a possible class of
compact objects contributing to the faint X-ray source population
in the Galaxy.  In the next section, the X-ray data on millisecond
pulsars is collected, illustrating the observational correlations
of their X-ray properties.  The theoretical basis for these
correlations are described in \S 3 within the framework of a
simple model, based on the work by Chevalier (2000) for 
the wind nebulae surrounding young pulsars.  In addition, 
the emission from an intrabinary shock for the case in which 
the relativistic pulsar wind interacts with a companion star is 
also discussed. In \S 4, the model is applied to the millisecond 
pulsar B 1957+20 and is used to estimate the pulsar wind contribution 
to the non-pulsed X-ray luminosity in the millisecond pulsars B1937+21 
and J 0218+4232.  We also discuss the possible 
relevance of the emission from the wind nebulae and surrounding 
environment of millisecond pulsars in the soft X-ray 
transient SAX J1808.4-3658, the Galactic center, and in the globular 
clusters 47 Tuc and M28.  Finally, we summarize and discuss the 
implications of the results of our study in the last section.

\section{OBSERVATIONAL X-RAY PROPERTIES OF MILLISECOND PULSARS}

The ROSAT results from the early studies of Becker \& Tr\"umper
(1993, 1999) suggested that the X-ray radiation from millisecond
pulsars is composed of non-thermal and thermal components.  The
thermal emission was attributed to the heated polar caps and the
non-pulsed non-thermal emission to a pulsar wind or a plerion
(Becker \& Tr\"umper 1993). Observations with the ASCA satellite
at energies higher than those accessible by the ROSAT satellite
have also revealed the presence of X-ray pulses (Saito et al.
1997b). More recently, Mineo et al. (2000) and Becker \&
Aschenbach (2002) detected both the pulsed and non-pulsed
non-thermal emission components of additional millisecond pulsars
based on {\em BeppoSAX, Chandra} and {\em XMM-Newton} observations. 
Direct evidence of the spatial X-ray distribution was provided 
by Stappers et al. (2003) in their important discovery of the nebula 
surrounding PSR B1957+20.

In Table 1, the luminosities and spectral properties of 8 millisecond 
pulsars in the  Galactic field are collected from recent observations 
based on X-ray data obtained from the {\em ASCA, RXTE, BeppoSAX, Chandra} 
and {\em XMM-Newton} satellites. For 2 of these pulsars, the pulsed 
and/or non-pulsed emission components were separately analyzed.  As in 
our previous study (Cheng, Taam, \& Wang 2004), we assume that
the pulsed component arises from the pulsar magnetosphere and that
the non-thermal non-pulsed component is due to a pulsar wind
nebula. 

The relation between the total X-ray luminosity and the spin down
power of the 8 millisecond pulsars, as obtained from the
literature, is illustrated in Fig. 1. A correlation between the total
X-ray luminosity in the energy range of 2-10 keV, $L_x^{tot }$, and 
the spin down power, $\dot E$, is seen to be present, and it is 
described by the best fit function $L_{\rm x} \propto \dot E^{1.39 
\pm 0.08}$ ergs s$^{-1}$, where we have approximated the errors in 
log($L_x$) to be $\pm$ 0.2.  This relation was first discovered 
by Seward \& Wang (1988) with Einstein data and others have also found 
similar correlations of normal and millisecond pulsars taken together 
(e.g. $\ddot{\rm {O}}$gelman 1995; Saito 1998; Cheng, Taam \& Wang 2004).  
Becker \& Tr\"umper (1997), using  ROSAT data in the energy range of 0.1-2.4 
keV, found a slightly different relation ($L_x \propto \dot E^{1.03 
\pm 0.08}$). In the energy range analyzed by Becker \& Tr\"umper (1997), the 
thermal component, which is expected to be emitted from the neutron star 
surface, can play a significant role. On the other hand, the thermal 
component should not significantly contribute in our selected energy range 
(2-10 keV). This suggests that there could be another component 
contributing to the harder X-rays. Cheng, Taam \& Wang (2004) have used the 
ASCA data of both normal and millisecond pulsars to analyze the relations 
between the $\dot E$ and the total X-ray luminosity, pulsed X-ray 
luminosity ($L_x^{pul}$), and non-pulsed X-ray luminosity ($L_x^{npul}$). 
The relations between $L_x^{tot}-\dot E$ and $L_x^{npul}- \dot E$ 
are found to be very similar, whereas the relation between $L_x^{pul}-\dot E$ 
is similar to the relation obtained by Becker \& Tr\"umper (1997). 
It was shown that the latter relation results from magnetospheric 
radiation, but that the non-pulsed radiation likely originates from the 
pulsar wind nebula (PWN). Since ASCA has a large collection area, which should 
cover the pulsar wind shock radiation region, a significant fraction of the 
total luminosity could be contributed by the PWN. Although 
we do not have a sufficient number of millisecond pulsars to carry out the 
non-pulsed X-ray luminosity analysis, we hypothesize that the pulsar wind 
nebula is a significant contributor to the total luminosity in 2-10 keV range.

The measured photon power law indices, $\Gamma$, of the total X-ray emission 
are found in the range from $\sim 0.6 - 3$ (see Table 1). For the 
millisecond pulsars in which the photon index of the pulsed and non-pulsed 
components were determined, it appears that the spectral indices of their
non-pulsed emission range from 3.3 for B1937+21 to 1.17 for J0218+4232.

\section{THEORETICAL PROPERTIES OF X-RAY EMISSION FROM MILLISECOND PULSARS}

The X-ray emission from a pulsar can arise from various processes
on the neutron star surface or in its vicinity.  For example,
thermal processes can occur on its surface and polar cap regions,
synchrotron emission processes within its magnetosphere and
synchrotron processes in the shocked region resulting from the
interaction of a relativistic wind with a binary companion to
the pulsar or the surrounding interstellar medium. It should
be pointed out that curvature radiation and the inverse Compton 
processes can also contribute to the radiation from the 
pulsar magnetosphere and pulsar wind nebula. However, these 
two processes can only generate $\gamma$-rays. For example, 
in the case of the intrabinary shock, it has been shown that although the
inverse Compton scattering can be the dominant cooling process, it will
produce a Maxwellian spectrum with a peak energy at several tens of GeV
(Tavani, Arons \& Kaspi 1994; Tavani \& Arons 1997). In this paper we 
focus on the X-ray emission, and, therefore we will ignore these two processes.  
In the following, we discuss the various emission processes in turn.

\subsection{Emission inside the light cylinder}

As thermal X-ray emission from the entire neutron star surface is
not expected to be appreciable due to the age of the pulsar, the thermal
X-ray emission from the neutron star surface is likely to be
limited to a polar cap region. For simplicity, we assume that the thermal
X-ray luminosity is given as $L_{\rm X,th}\simeq \dot N V_{\rm
gap}$.  Here, $\dot N\sim 1.35\times 10^{30} B_{12}P^{-2}{\rm e\
s^{-1}}$ (Goldreich \& Julian 1969), and $V_{\rm gap}$ is
the potential of the polar gap. Jones (1980) showed that the photoejection
of the most tightly bound electrons of ions can limit the potential of the polar gap
to $V_{\rm gap}=\gamma(A/Z)10^9$ V where $A/Z \sim 2$ and 
$\gamma$ is the Lorentz factor of ions.
Cheng \& Taam (2003) argue that if higher
order magnetic fields are present on the surface of neutron star 
$\gamma \sim 20$. For the typical parameters
of millisecond pulsars in the Galaxy (e.g. $P=3$ ms, $B=2\times
10^8$ G), $L_{\rm X,th}\sim 10^{30}{\rm erg\ s^{-1}}$.

In contrast to the thermal processes taking place on the neutron
star surface, non-thermal X-rays can be produced in its immediate
vicinity as a result of synchrotron emission of electron-positron
pairs produced by an electromagnetic cascade near the neutron star
surface (see Cheng, Gil \& Zhang 1998; Cheng \& Zhang 1999; Zhang \& Cheng 2003). 
The model  predicts a luminosity $\sim 2.5\times 10^{-9}\dot{E}^{1.15}$ 
(Cheng \& Zhang 1999).  The conversion efficiency of spin down power to 
photon luminosity in the energy range from 0.1 keV to 1 MeV is $\sim 3 
\times 10^{-4}$.

\subsection{Pulsar wind nebulae in the interstellar medium}
The non-thermal X-ray radiation can also be produced as a result
of synchrotron emission processes in a region affected by the
interaction of the pulsar's relativistic wind with the
interstellar medium. The resulting interaction can take the form
of a termination shock. A detailed description of the 
interaction between a pulsar and its nebula, as applied to the Crab 
Nebula, was first outlined in a seminal paper by Rees and Gunn (1974). 
Theoretical studies on the X-ray emission from PWN have remained as 
an important topic due to the increased quality of observed data. 
Blondin, Chevalier \& Frierson (2001) have used one- and two-dimensional
two fluid models to simulate the PWN in evolved young supernova remnants. 
This model can explain some chaotic and asymmetric appearances of young PWN 
such as the Vela Nebula. Bucciantini (2002) has used a 2D hydrodynamic 
model to estimate the opacity of the bow-shock to penetration of ISM neutral
hydrogen, which can significantly affect the size, shape, velocity and 
brightness distribution of the nebula. A magnetohydrodynamical model 
was developed by van der Swaluw (2003) to simulate the expansion of PWN, showing 
that the PWN is elongated as a result of the dynamical effect of toroidal 
magnetic fields. Finally, Komissarov \& Lyubarksi (2003) have used a 
2D relativistic magnetohydrodynamical model to simulate the peculier 
jet-torus structure of the Crab nebula.  For a recent discussion of the emission
characteristics of the shocked region, see Cheng et al. (2004).
In this subsection, we will use a simple spherically symmetric (1D) model 
similar to that suggested by Chevalier (2000) to estimate the non-thermal 
radiation from the PWN of MSP. In general, 2D or 3D models are necessary 
to provide a detailed theoretical explanation of the emission 
morphology from young pulsars (Gaensler 2004).  However, the proper motion 
of MSPs are typically much slower than canonical high magnetic field pulsars.
The PWN of MSP may be older and expanding more slowly (subsonically).  
Furthermore, we can estimate the distance from the neutron star where 
the charged particles in the pulsar wind can
break away from the pulsar magnetic field by requiring that the Larmor radius
of the particle becomes larger than the curvature radius of the magnetic field. 
This requirement leads to a break away radius at 
$r_{break}\sim \frac{qR^3B_s\Omega}{cE_p}$, where $B_s$ is the surface magnetic field,
$R$ is the stellar radius and $E_p\sim \gamma_w m_p c^2$ is the mean energy 
of the charged particle in the wind. Although the toroidal 
magnetic fields of young pulsars in PWN could indeed have a dynamical
influence on its structure, making the nebula elongated (e.g. van der 
Swaluw 2003), the magnetic field strengths of MSPs are much weaker and 
may not be as effective. From the above estimate, we can see that 
this is true if the mean Lorentz factor 
in the pulsar is a constant. Cheng, Taam and Wang (2004) have argued that 
the relation between the X-ray luminosity and the pulsar
spin-down power ($\dot{E}$) can be obtained if $E_p\propto \dot{E}^{1/2}$. 
Since $B_s\Omega^2\propto \dot{E}^{1/2}$, then $r_{break}\propto 1/\Omega$, 
providing additional evidence 
that the nebula of MSP is not as elongated as in the young pulsars.
Consequently, the bow shock created by MSP may be well approximated by a 1D model.

For isolated millisecond pulsars, the radius of the bow shock
(which is expected to be comparable to the size of the X-ray emitting region)
produced by the interaction of the pulsar wind with the
interstellar medium is obtained from a pressure balance condition,
leading to 
\beq R_s = ({\dot E}/2\pi \rho v_p^2 c)^{1/2}\sim
1.8\times 10^{16}{\dot E}_{34}^{1/2}n^{-1/2}v_{p,100}^{-1} {\rm cm}, \enq
where $v_{p,100}$ is the pulsar velocity in units of 100 km
s$^{-1}$, $\dot E_{34}$ is the spin down power of the pulsar in
units of $10^{34}$ ergs s$^{-1}$, and $n$ is the number density of
the interstellar medium in units of 1 cm$^{-3}$. We point out that 
the size of PWN is energy dependent since the cooling time in X-rays 
is shorter than in the radio regime.  Since it is very likely that
the synchrotron cooling time in X-ray energy range is shorter than the
flow time in the shock region, the X-ray PWN can be approximated
by Equation 1. On the other hand, the TeV PWN should be larger 
than that of X-ray PWN because when the relativistic electrons leave 
the shock region (while radiating at lower frequencies), they contribute 
to the TeV photon production via the inverse Compton scattering process.

In the shock wave, the energy can be stored in the proton (ion),
electron, and magnetic field components. The fractional energy
densities $\epsilon_p$, $\epsilon_e$, and $\epsilon_B$ correspond
to that of the protons, electrons, and magnetic field
respectively. We note that the protons do not contribute to the
X-ray emission through the synchrotron process. The energy
densities in the shock are not theoretically determined with any
certainty, however, Kennel \& Coroniti (1984) have argued that
only for ratios of magnetic to kinetic energy, defined as $\sigma_B
\sim \epsilon_B/(\epsilon_p+\epsilon_e)$, less than about 0.1 can
a significant fraction of total energy flux upstream be converted
into thermal energy downstream and, thereafter, into synchrotron
luminosity. In fitting to the observational data of the Crab
nebula, Kennel \& Coroniti (1984) found that $\sigma_B$ should  
lie in the range between 0.001 and 0.01.  deJager \& Harding (1992) 
have found that $\sigma_B =0.003$ gives the best fit for the EGRET and 
TeV data from the Crab nebula.  
Recently, Sefako \& deJager (2003) have found that the broadband spectrum of 
a PWN crucially depends on the magnetization parameter, spin-down 
power, distance, particle multiplicity and shock radius. In order to fit the broadband
spectrum of the Vela pulsar and PSR  B1706-44, the best fit value of $\sigma_B =0.1$ 
was found for these pulsars, both of which have very similar ages ($\sim 5\times 
10^3$years). This value is substantially different from that of the Crab nebula. 
Since the Crab nebula has been observed and theoretically studied more 
than any other pulsar wind nebula, we adopt $\sigma_B$=0.003 as the standard
reference  value, but we also consider other possible value of $\sigma_B$.
In addition, we also assume comparable particle kinetic
energy fractions (e.g, $\epsilon_p$ and $\epsilon_e \sim 0.5$)
throughout this study. Thus, the magnetic field in the emitting
region can be estimated as 
\beq B = (6\epsilon_B \dot E/R_s^2
c)^{1/2}\sim 5\times 10^{-6} (\epsilon_B/0.003)^{1/2}n^{1/2} v_{p,100} G.
\enq

Following the analysis of Chevalier (2000, see also Cheng et al.
2004), the X-ray properties of the pulsar wind can be
theoretically calculated. In particular, we take the postshock
electron energy distribution for a strong relativistic shock as
$N(\gamma)\propto \gamma^{-p}$ for $\gamma > \gamma_m$, where
$\gamma_m = {p-2 \over p-1}\epsilon_e \gamma_w$, and $\gamma_w$
is the Lorentz factor of the relativistic pulsar wind, which is
assumed to be $10^6$ throughout this study.

The determination of the X-ray luminosity is dependent on the
cooling frequency, $\nu_c$, given as $\nu_c= {e\over 2\pi
m_ecB^3}({6\pi m_ec\over\sigma_T t})^2$, where $t$ is a
characteristic timescale of the nebula estimated as the flow
timescale in a characteristic radiation region, i.e, $t\sim
R_s/v_p$. For $\nu_X > \nu_c$ (denoted as fast cooling), the observed
luminosity of radiating particles per unit frequency is given as
(Chevalier 2000) $$L_\nu = {1\over 2}({p-2\over
p-1})^{p-1}({6e^2\over 4\pi^2 m_e c^3})^{(p-2)/4}
\epsilon_e^{p-1}\epsilon_B^{(p-2)/4} \gamma_w^{p-2}$$ \beq R_s^
{-(p-2)/2}{\dot E}^{(p+2)/4}\nu^{-p/2}. \enq Here, we estimate the 
total X-ray luminosity as $\nu L_\nu$ for simplicity.  Taking typical 
parameters for millisecond pulsars ($P=3$ ms, $B=2\times 10^8$ G, 
$\dot E \sim 10^{34}$ ergs s$^{-1}$, $v_p \sim 180$ km s$^{-1}$ and 
$n \sim 1\rm cm^{-3}$, $\epsilon_e \sim 0.5,
\epsilon_B \sim 0.003, \gamma_w \sim 10^6$, and $p=2.2$ (we
take the possible optimum value based on theoretical
investigations of highly relativistic shock fronts in the limit of
high Lorentz factors, see Bednarz \& Ostrowski 1998, Lemoine \&
Pelletier 2003), the X-ray luminosity ($\nu \sim 10^{18}$Hz) is
$\nu L_\nu \sim 4\times 10^{30}{\rm ergs\ s^{-1}}$.  
In order to explain the observed relation between the X-ray luminosity and
the spin-down power, we (Cheng, Taam \& Wang 2004) have argued that 
$\gamma_w\sim 5\times 10^3B_{12}/P^2 \sim 2\times 10^5 (B/3\times 10^8G)/(P/3ms)^2$. 
For $\gamma_w=2\times 10^5$ and $\epsilon_B \sim 0.1$, 
$\nu L_\nu$ will decrease slightly to $ \sim 3.3\times 10^{30}{\rm ergs\ s^{-1}}$.
If we decrease $p$ to slightly larger values than $\sim$2, 
the luminosity reaches a level of $\sim 4\times 10^{31}{\rm ergs\ 
s^{-1}}$. In this regime, the efficiency of conversion of spin down 
power to X-ray luminosity can be as high as $\sim 4\times 10^{-3}$, a 
value which is much greater than that associated
with the pulsed emission from the magnetosphere (cf. \S 3.1).  As is well 
known, the photon index in the fast cooling regime is given 
by $\Gamma=(p+2)/2$. Generally, $p$ can vary from 2
to 3, and the pulsar wind nebulae could produce non-thermal
spectra with photon indices in the range from 2 to 2.5. On
the other hand, if the cooling frequency is larger than the
frequency $\nu_X$ (denoted as slow cooling), the observed luminosity per
frequency $L'_\nu \sim (\nu/\nu_c)^{1/2}L_\nu \propto
\nu^{-(p-1)/2}$. In this regime, the X-ray luminosity will be
lower than that in the fast cooling regime, and the observed
photon index is $\Gamma =(p+1)/2$ (Chevalier 2000). In this case, 
the pulsar wind nebulae could produce non-thermal spectra with photon 
indices ranging between 1.5 and 2, corresponding to conversion
efficiencies ranging from about $2\times 10^{-6}$ to $9\times 10^{-5}$ for 
the parameters used in the fast cooling case (see above). 

In principle, the theoretical model can be tested using the
observed X-ray luminosity, the size of the X-ray emission region
($\sim R_s$), and Eq. 3 to solve for the value of $p$ to determine
if it is within the theoretical allowed range, thus providing a
consistent estimate of $\Gamma$ in comparison with the observed  
spectral index. However, a spectral index has to be assumed to
infer the observed X-ray luminosity. In order words, $\Gamma$ and
$L_x$ are not independent. Furthermore, the errors in $L_x$ and
$R_s$ do not provide an accurate estimate of $p$. Therefore, it is
more practical to choose a possible value of $p$ with the constraint
of theories (e.g. Bednarz \& Ostrowski 1998, Lemoine \& Pelletier
2003) to estimate $L_x$ and $\Gamma$ for comparison with those
observed. We remark that since shock acceleration is a non-linear 
process, it would not be surprising that
various observations lead to a range of $p$. In the following
applications, we will treat $p$ as a parameter but with the
constraint that it should only vary between 2 and 3 as suggested
by theoretical studies (e.g. Bednarz \& Ostrowski 1998, Lemoine \&
Pelletier 2003).

\subsection{Emission from the intrabinary shock}

Non-thermal radiation can also arise  from a shock
wave resulting from the interaction between the pulsar wind and 
the outflowing matter from the companion star (Arons \& Tavani 1993, 
Stella et al. 1994, Tavani \& Arons 1997).  In this context, the 
neutron star is at a distance, $D$, from the surface of its binary 
companion given by $D=R+R_s$, where $R_s$ is the shock wave termination 
radius as measured from the neutron star surface and $R$ is the distance 
of the shock from the surface of the companion star. For millisecond 
pulsars in binary systems with low mass companions, we adopt  
an orbital separation $\sim 2 \times 10^{11}$ cm. If we assume that mass is lost 
isotropically from the companion star, the loss rate is given as $\dot M 
=4\pi \rho (R_* + R)^2 v_w$, where $v_w$ is its  outflow velocity and $\rho$ 
is the density at distance $R$ from the stellar surface and $R_*$ is the radius 
of the companion.  The dynamic outflow 
pressure given by $P_w(R)=\rho(R)v_w(R)^2$ can be expressed as 
\beq P_w(R)= {\dot M v_w \over 4 \pi (R_*+R)^2}. \enq  
The termination radius of the pulsar wind is given by the pressure
balance between the pulsar and companion outflow  and can be expressed as 
\beq \left({R_s\over R_* + R} \right)^2 = {\dot{E} \over \dot{M} v_w c}. \enq

In order to estimate the shock radius, $\dot{M}$ and $v_w$ must be 
determined. Unlike the case of the progenitor of millisecond pulsars, i.e., 
the low mass X-ray binaries, where $\dot{M}$ can be estimated from the 
X-ray accretion luminosity, the mass loss rate from the neutron star companion 
likely results from the evaporation by a pulsar wind (Ruderman, Shaham \& 
Tavani 1989). If the stellar wind is a consequence of evaporation by the 
pulsar wind, the balance between the stellar wind pressure and the pulsar 
wind pressure lies near the position of the companion, implying $R_s \sim D$.
In taking PSR B1957+20 as an example, we find that $R_s > 0.6 (R_* + D)$
under the assumption that the wind velocity is comparable to the escape speed from
the companion and the mass loss rate is  $\lesssim  10^{17}$ g s$^{-1}$ (see 
Stappers et al. 2003).  

The cooling frequency $\nu_c$ is given as $\nu_c= {e\over 2\pi m_ecB^3}
({6\pi m_ec\over\sigma_T t_f})^2$, where $t_f \sim \sqrt{3} R_s/c$ is 
the dynamical flow time and the magnetic field strength at the termination 
radius is estimated by $B = (6\epsilon_B \dot E/R_s^2 c)^{1/2}$.
Taking $\dot E \sim 10^{35}$, $\epsilon_B \sim 0.003$, and $R_s \sim 
10^{11}$cm, we obtain $\nu_c \sim 3\times 10^{21}$Hz.  Since the 
cooling frequency is much 
larger than the frequency $\nu_x \sim 10^{18}$ Hz, the observed luminosity per 
frequency $L_\nu \propto \nu^{-(p-1)/2}$.  Based on the work by 
Chevalier (2000), the total X-ray 
luminosity radiated in the intrabinary shock wave for a solid angle of 
$\Omega$ toward the pulsar is estimated as 

$$\nu L_\nu ={\Omega\over 4\pi}{\sigma_T 6^{(p-3)/4}(p-2)^{p-1}\over 
2\pi m_ec^{(p+5)/4}(p-1)^{p-2}}({e\over 2\pi m_ec})^{(p-3)/2}\epsilon_e^{p-1}\epsilon_B^{(p+1)/4} $$ 
\beq R_s^{-(p+1)/2}\gamma_w^{p-2} t_f \dot{E}^{(p+5)/4}\nu^{-(p-3)/2},
\enq 
in the X-ray energy band (2-10 keV).  It is interesting to note that if 
we approximate $t_f \sim \sqrt{3} R_s/c$ and  $R_s =D$, the above equation 
can be rewritten as 

\beq \nu L_{\nu} = 5 \times 10^{30} \alpha(p) f_{0.1} D_{11}^{-(p-1)/2} 
\dot{E}_{35}^{(p+5)/4} {\rm ergs } \ {\rm s}^{-1},
\enq 
for $\nu =10^{18}$Hz, $\gamma_w = 10^6$, $\epsilon_e =0.5$,  $\epsilon_B =0.003$ and $\alpha(p)$ is a 
function of $p$, which only  varies from 1 to 2.6 as p increases from 2.2 to 
2.6. For simplicity, 
we may assume it is a constant of order of unity. Here, $f=\Omega/4\pi$, 
$D_{11}= D/10^{11}$cm and $\dot{E}_{35} = \dot{E}/10^{35}$ ergs s$^{-1}$.
We note that Tavani \& Arons (1997) have estimated that $\epsilon_B 
\sim 0.02$ in the intrabinary shock region in order to explain the X-ray 
emission from PSR 1259-63.  For this value, the coefficient in Equation 7 
will change to $3\times 10^{31}$.

Thus, the emission from the intrabinary shock is expected to be in 
the slow cooling regime characterized by a photon index, $\Gamma=(p+1)/2$, 
while the emission from the pulsar interaction with the interstellar 
medium can be either in the slow or fast cooling regime.

\section {APPLICATIONS}

\subsection{Pulsar Wind Nebula Source}

\subsubsection{PSR B1957+20}

In a recent study by Stappers et al. (2003) an X-ray nebula in the
vicinity of the millisecond pulsar B1957+20 was detected. A bright
X-ray source in the field was found to be coincident with the
pulsar position, and a tail of X-ray emission was seen extending
from the pulsar to the northeast by at least 16" (corresponding to
$\sim 4\times 10^{17}$ cm at the distance of 1.5 kpc) at a
position angle opposite to the pulsar's direction of motion. Taken
together with the existence of a previously identified H$\alpha$
bow shock structure (Kulkarni \& Hester 1988) enveloping this
tail, PSR B1957+20 is a prime example of an X-ray bow shock nebula
in a pulsar binary system.  PSR B1957+20 is the second fastest
spinning pulsar known, with a rotation period of 1.6 ms and a
rotational spin-down luminosity of $\simeq 10^{35}{\rm erg\
s^{-1}}$ (Fruchter et al. 1988, Toscano et al. 1999). It is a
member of a binary system ($P = 9.16$ hour) with a low mass
companion star separated by a distance of $1.5\times 10^{11}$ cm.
Since PSR B1957+20 has a non-thermal X-ray spectrum ($\Gamma=1.9
\pm 0.5$), it is likely that the pulsar wind interaction 
with the companion and the interstellar medium, and the pulsar's
magnetosphere contribute to the unresolved emission. Stappers et
al. (2003) argue that the unresolved X-ray emission originates
from the interaction between the pulsar wind
and the induced stellar wind of its companion, and that the
resolved emission in the tail is a result of the interaction
between the pulsar wind and the interstellar medium.

The tail is characterized by a length $l \gtrsim 4\times 10^{17}$ cm
and an X-ray luminosity of $\sim 2.4 \times 10^{30}$ ergs
s$^{-1}$.  We interpret its length in terms of the distance
traversed by the pulsar within the electron synchrotron cooling
timescale,i.e. $l \sim v_p t_c$ where $v_p$ is the proper motion 
velocity of the pulsar. The magnetic field strength in the region 
surrounding the pulsar can be roughly estimated by assuming a 
compression of the interstellar magnetic field ($\sim 2 - 6 \mu$G; 
see Beck et al. 2003). For a field compression in the termination 
shock of a factor of 3 (Kennel \& Coroniti 1984), the inferred  
field is $\sim 6 - 18 \mu$G.  The synchrotron cooling time in the
X-ray band is $\tau_c=6\pi m_ec/\gamma \sigma_T B^2 \sim 10^8
B_{mG}^{-3/2} (h\nu_X/ {\rm keV})^{-1/2}$ s, yielding $\tau_c$ 
in the range of 1400-7100 yrs.  Since the length of the tail
is energy dependent, we have chosen a representative X-ray energy 
at KeV in order to compare with the observations. The pulsar moves through
the interstellar medium at $220 {\rm km\ s^{-1}}$ (Arzoumanian et
al. 1994), and the inferred tail length is $\sim 9 \times 10^{17} - 
4.6 \times 10^{18}$ cm, suggesting that the higher inferred 
magnetic field is more appropriate. Alternatively, equating 
the observed tail length to the product of the synchrotron cooling 
timescale and the pulsar velocity leads to $B \sim 30 \mu$G or 
a factor of 1.67 higher than the previously estimated value. 

For a number density $\sim 1$ cm$^{-3}$, consistent with the
hydrogen column density inferred from the X-ray observations
(Stappers et al.  2003), the termination shock radius of PSR
B1957+20 is estimated to be $\sim 2.5\times 10 ^{16}$ cm, which
corresponds to an angular extent of $\sim$ 1".  Given 
$B_{mG} = 0.03$ (see above), we
estimate that the cooling frequency is lower than the X-ray
frequency.  Based on the results in \S 3.2 the spectral index of the
nebula component is then $\Gamma=(p+2)/2$.  Taking $p=2.3$,
$\epsilon_B\sim 0.003, \epsilon_e\sim 0.5$ and $\gamma_w =
10^6$, and $R_s\sim 2.5\times 10^{16}$ cm, the predicted nebula
luminosity (0.5-7 keV) is $\sim 5\times 10^{30}{\rm erg\ s^{-1}}$.
If we choose $\epsilon_B\sim 0.1$ and $\gamma_w =5\times 
10^5$, then the luminosity increases to
$7\times 10^{30}{\rm erg\ s^{-1}}$. This value 
is higher than the observed luminosity in the tail by a factor of
about 3, suggesting that some of this emission may contribute
to the unresolved emission (see below). 

The X-ray luminosity of the unresolved source in an aperture
radius 1.5" is $\sim 1.6\times 10^{31}{\rm erg\ s^{-1}}$, and 
Stappers et al. (2003) suggest that it is produced in an intrabinary 
shock between the pulsar wind and that of the companion star.
According to the estimates given in \S 3.3, the X-ray luminosity 
(0.5-7 keV) from the intrabinary shock, which contributes to the unresolved
component, is about $\sim 5 \times 10^{30}{\rm erg\ s^{-1}}$ 
assuming $p=2.3, \epsilon_B\sim 0.003, \epsilon_e\sim 0.5, \gamma_w = 10^6$, 
and $\Omega=1$. Again choosing $\epsilon_B\sim 0.1, \gamma_w = 5\times 10^5$,
the X-ray luminosity from the intrabinary shock increases significantly to
$\sim 6 \times 10^{31}{\rm erg\ s^{-1}}$. In order to fit the observation,
$\epsilon_B \sim 0.02$. This value is consistent with the value obtained by
Tavani \& Arons (1997) for PSR1259-63, where the X-rays and $\gamma$-rays also 
result from intrabinary shock.

>From the above estimates, it is evident that the pulsar wind 
nebula could contribute to the flux of both 
the unresolved X-ray source as well as to the X-ray tail since 
the termination radius ($R_s\sim 2.5\times 10^{16}$ cm) subtends 
an angle comparable to the {\em Chandra} angular resolution. 
In this case, the pulsar wind interaction with the 
interstellar medium would contribute  significantly to the flux 
of the unresolved X-ray source.  With the contribution 
from the intrabinary shock interaction, the total predicted nebula 
X-ray luminosity would amount to $\sim 1.4\times 10^{31}{\rm 
erg\ s^{-1}}$, which is consistent with the observed value.

Becker \&  Tr\"umper (1997,1999) have suggested that the X-ray
luminosity in the range of 0.2-2.4 keV of rotation powered pulsars 
should satisfy $L_x \sim 10^{-3} \dot{E}$. This relation predicts  
that the X-ray luminosity in the 2-10 keV energy range to be a factor 
of $5_{+6}^{-3}$ larger than observed assuming a photon index $\Gamma 
= 1.9\pm 0.5$. Cheng \& Taam (2003) 
have argued that if there are higher order magnetic fields present 
on the surface of millisecond pulsars, the non-thermal X-rays as well as
gamma-rays will be suppressed. In this case, the thermal X-ray
luminosity from the polar cap heating will be reduced to $\sim 3
\times 10^{30}$ ergs s$^{-1}$ as suggested in \S 3.1.

\subsubsection{PSR B1937+21 and J0218+4232}

Of the millisecond pulsars listed in Table 1, only PSR B1937+21
and J0218+4232 have measured power law photon indices for 
both the pulsed and non-pulsed emission components, partly because 
they have a larger spin down power and higher X-ray luminosity than 
the other millisecond pulsars. Although pulsar wind nebulae have
not been imaged in these two millisecond pulsars, it is possible
that the non-thermal, non-pulsed components are produced within
such regions. In this subsection, we explore the consequences of
pulsar wind nebulae surrounding these millisecond pulsars as the
candidate sites for this X-ray emission component.

The velocities of PSR B1937+21 and J0218+4232 are still unknown or 
quite uncertain. Since the average birth velocities of observed 
millisecond pulsars are $\sim 130 {\rm km\ s^{-1}}$ (Lyne et al. 1998), 
we  adopt this value as their typical velocity. Assuming that
the surrounding medium has a density of $\sim 1$ cm$^{-3}$ and
using the spin down power listed in Table 1, the termination
radius of the bow shocks in the hypothesized nebulae are
$\sim 2\times 10^{17}$ and $\sim 10^{17}$ cm for B1937+21 and 
J0218+4232 respectively.  Taking a conservative value of 
$\epsilon_B \sim 0.003$, the magnetic field is estimated to 
be $\sim 10^{-5}$ G, and the cooling frequencies of the
nebulae are given as $\sim 3.6 \times 10^{19}$ Hz for 
B1937+21 and $1.4 \times 10^{20}$ Hz for J0218+4232. Since $\nu_c
> \nu_X$, we find the nebulae would be in the slow cooling
regime.  As a result, the simple one-zone model outlined in \S 3
indicates that their photon spectral indices, $\Gamma=(p+1)/2$,
are generally smaller than 2.
The observed non-pulsed photon index of PSR B1937+21 is $3.3\pm 0.5$, which 
should be more consistent with the fast cooling region. 
In this case if we take $\epsilon_B \sim 0.1$,
the cooling frequencies of the
nebulae are given as $\sim 1.8 \times 10^{17}$ Hz for 
B1937+21 and $7.2 \times 10^{17}$ Hz for J0218+4232 respectively. This
means $\nu_c < \nu_X$ and these systems are in fast cooling regime.
Furthermore, its
non-pulsed component is likely contaminated by the thermal component 
and the actual photon index for non-thermal emission may be closer to 2, 
which implies $p\sim 2.3$.  In this case, the predicted X-ray luminosity and
the observed data would be consistent.
On the other hand, the observed non-pulsed 
photon index of PSR B0218+4232 is $1.17\pm 0.37$, which indeed is in 
slow cooling region.  If we adopt the typical value of $p=2.3$, the 
predicted X-ray luminosity is $\sim 7\times 10^{31}{\rm ergs\ s^{-1}}$, 
which is higher than the observed value by a factor of 2.5. Since 
estimates of the distance to J0218+4232 could be as large as 5.7 kpc 
(Cordes \& Lazio 2002), the discrepancy may not be as large as implied.
Given the simple model that we have adopted and the sensitivity of 
the luminosity to the power law index of the electron 
energy distribution, it is encouraging that the luminosities 
from the one zone model are of the same order as observed.

\subsubsection{PSR J2124-3358}

PSR J2124-3358 was discovered during the Parkes 346 MHz survey by Bailes et al. (1997), 
and X-ray emission was subsequently detected in ROSAT HRI data by Becker \& Tr\"umper 
(1998, 1999). This pulsar has a period of 4.93ms and is characterized by a surface 
magnetic field of $2.6 \times 10^8$G, which gives a spin-down power $4.4\times 10^{33}$
erg s$^{-1}$. The estimated distance, based on the dispersion measure, is 250pc, and 
the proper velocity is 58 km s$^{-1}$ (Manchester et al. 2005). Very recently  
the X-ray luminosity associated with a PWN was reported by Hui \& Becker (2005). The 
observed X-ray luminosity associated with the nebula is estimated to be $\sim 10^{29}$ 
ergs s$^{-1}$, and the spectrum was fitted with a photon power index of $2.2\pm 0.3$. 
The X-ray emission extends from the pulsar to the northwest by $\sim$1 arcmin, which 
corresponds to a linear scale of $\sim 2\times 10^{17}$ cm. 
The time scale for the passage of the pulsar 
over this length is about 1,100 years. Assuming that such an X-ray tail results from 
the finite synchrotron life time effect, we can estimate the magnetic field in the X-ray 
emission region, which gives B$\sim 20\mu$G.  Although this value is larger than the 
typical magnetic field in the interstellar medium of 2-6$\mu$G (Beck et al. 2003), 
it is consistent with the fact that the magnetic field in the termination shock 
can be enhanced by a typical factor of 3 (Kennel \& Coroniti 1984). With this magnetic
field we estimate a cooling frequency of $\sim 2\times 10^{17}$Hz, which suggests that the X-ray 
emission is in fast cooling regime. The observed photon index suggests that $p\sim 2.3$.
For $\epsilon_B\sim 0.003, \epsilon_e\sim 0.5$ and $\gamma_w = 10^6$, the estimated 
X-ray luminosity is $2\times 10^{29}$ ergs s$^{-1}$, while for $\epsilon_B\sim 0.1$ and 
$\gamma_w = 2\times 10^5$, it is $1.6\times 10^{29}$ ergs s$^{-1}$ indicating that 
the X-ray luminosity is not sensitive to these chosen parameters. We note that these
luminosity estimates are consistent with the observed data (Hui \& Becker 2005).

\subsection {POTENTIAL PULSAR WIND NEBULA SOURCES}

\subsubsection{SAX J1808.4-3658}

SAX J1808.4-3658 was the first low-mass X-ray binary system
discovered in which the accreting pulsar exhibited coherent
pulsations at a spin period ($\sim 2.5$ ms) comparable to that
observed in millisecond radio pulsars (Wijnands \& van der Klis
1998).  The source is a soft X-ray transient (SXRT) which reaches
a maximum X-ray luminosity of $\sim 2\times 10^{36} {\rm ergs\
s^{-1}}$ (for a distance of 2.5 kpc; in 't Zand et al. 2001). {\em
XMM-Newton} observations of SAX  J1808.4-3658 were obtained during
its quiescent state, revealing an unabsorbed luminosity (0.5-10
keV) of $5\times 10^{31}{\rm erg\ s^{-1}}$, and a spectrum which
is characterized by a hard power law with photon index $\Gamma
\sim 1.5\pm 0.5$ and a minor contribution ($\leq 10\%$) from a
soft blackbody component (Campana \& Stella 2003). SAX 1808.4-3658
is distinguished from other SXRTs in quiescence by the fact that
the power law component is dominant (Campana \& Stella 2003).

Jonker et al. (2004) point out that the relative contribution of
the hard component to the total flux from neutron stars in SXT
systems in the quiescent state may be correlated with the
quiescent state X-ray luminosity level with a greater contribution
at lower quiescent X-ray luminosities.  However, the power law
component for Aql X-1 is dominant, even though the quiescent
luminosity is $\sim 2-4 \times 10^{33}$ ergs s$^{-1}$, suggesting
that the origin of the power law component may differ between
sources at high and low quiescent X-ray luminosities (Jonker et
al. 2004). Although the origin of this emission is unknown, it is
possible that the power law component for more luminous quiescent
sources arises, for example, from accretion onto the neutron star
whereas the power law component for low luminosity quiescent
sources may arise from the activation of the pulsar mechanism when
the accretion rate is significantly reduced (see Campana et al.
1998). If the pulsar mechanism is operative and a relativistic
wind is established, contributions to the luminosity and spectrum 
could involve the interaction between the pulsar wind with the interstellar
medimum and with a stellar wind of the companion star, the latter analogous
to that discussed for B 1957+20 and for PSR J0024-7204W in 47 Tuc (see 
below; Bogdanov, Grindlay, \& van den Berg 2005). Indirect evidence for 
the possible existence of a relativistic wind in SAX J1808.4-3658 is provided
by the requirement of an additional energy source to explain
the optical emission during the quiescent state (see Campana et
al. 2004).  The inferred magnetic field strength of SAX
J1808.4-3658 is in the range of $\sim 10^{8-9}$ G range (see
Psaltis \& Chakrabarty 1999), leading to a pulsar spin down power
of $\dot E \sim 1.8 \times 10^{31}B_{12}^2 P^{-4} {\rm ergs\ s^{-1}} \sim
6\times 10^{34-36} {\rm erg\ s^{-1}}$. If the efficiency of
conversion of spin-down power to X-ray luminosity is $\sim
10^{-5}-10^{-4}$, in the range found for pulsars with small values
of $\Gamma$ (see Cheng et al. 2004), the luminosity level due to
the interaction of the pulsar wind with the interstellar medium
could be consistent with that observed.  Such an interpretation
would suggest that the pulsar in SAX 1808.4-3658 is in the slow
cooling regime. A measurement of the space velocity of the system
would provide an estimate of the cooling frequency, yielding an
independent check on the consistency of this interpretation. 

\subsubsection{Galactic Center}

Deep {\em Chandra} X-ray surveys have led to the discovery of a
number of point-like and tail-like sources in the region of the
Galactic center (Wang et al. 2002; Muno et al. 2003). Since the
resolution limit of the {\em Chandra} observations is $\sim 1''$,
the tail-like sources are characterized by length scales exceeding
$10^{17}$ cm. Such tails may be similar to that described for the
millisecond pulsar B 1957+20 (see \S 4.1.1). Pulsar proper
motion velocities $v_p\gtrsim 100 {\rm km\ s^{-1}}$ and magnetic
field strength in the interstellar medium of B$ \lesssim 0.05$ mG,
comparable to the field in the Galactic center (see Morris 1994),
would be required to resolve such tail-like sources given the
resolution limit of the {\em Chandra} observations. For these
parameters, we find $\nu_c<\nu_X$, so millisecond pulsars will
have relatively higher luminosities and steeper spectra
$\Gamma\sim 2-2.5$ than if $\nu_c
> \nu_X$ (see \S 3). The photon power law index from such wind
nebulae is consistent with that inferred for the observed extended
sources (i.e., $\Gamma \sim 2$; see Lu, Wang, \& Lang 2003).  A
particularly unusual source was reported by Wang, Lu, \& Lang
(2002) who found an elongated feature G0.13-0.11 with a high X-ray
luminosity $\sim 10^{33} {\rm ergs\ s^{-1}}$ and long tail
$\gtrsim 10^{18}$ cm. It had been noted that a large radio
filament or arc is coincident with the X-ray emission region, and
Wang et al. (2002) suggested that this source is a young pulsar
wind nebula.  Although this interpretation is possible for such a
special source, it is also possible that a millisecond pulsar
interpretation is viable if the spin down power of the millisecond
pulsar is the order of $10^{35}{\rm ergs\ s^{-1}}$ and magnetic
field strengths in the surrounding medium of $B \sim 10 \mu$G. We
note that at least four millisecond pulsars have a spin down power
equal to or larger than this level (see Becker \& Aschenbach
2002). For slowly moving millisecond pulsars, the sources would
appear point-like. The photon
indices characterizing the spectra can lie in the range from 1.5
to 2.5 with the X-ray luminosities as high as $10^{31}-10^{32}{\rm
erg\ s^{-1}}$.

Since the {\em Chandra} resolution limit at the Galactic center is
about $10^{17}$ cm, which is larger than the X-ray emission radius
of a typical pulsar wind nebula, the X-ray emission of such
sources in the {\em Chandra} fields should include the additional
contributions from both the thermal component (due to emission
from the polar cap on the neutron star surface) and non-thermal
component (from the magnetosphere).  Since the thermal component
is soft (kT $< 1$ keV) and absorbed by interstellar gas for sources at the
Galactic center, only the non-thermal components associated with
the processes in the neutron star magnetosphere and the pulsar
wind nebula will contribute to the observed X-ray emission. We
note, however, that the luminosity of the non-thermal X-rays
produced as a result of synchrotron emission of electron-positron
pairs in the pulsar magnetosphere is not a significant contributor
to the total X-ray luminosity since the conversion
efficiency is $3 \times 10^{-4}$ (see \S 3).  Hence, the pulsar
wind nebula is likely to be the primary site of the "non-thermal"
X-ray emission. On the other hand, if the 
observed spectrum is poorly determined, hard "thermal" X-rays emitted from
the polar cap 
resulting from either polar cap heating of curvature radiation pairs or inverse Compton pairs 
(Harding \& Muslimov 2002)
can not be differentiated from the "non-thermal" hard X-rays. 

Recently, Park et al. (2004) have pointed out the existence of a
low density ($\sim 0.1$ cm$^{-3}$), hot gaseous component in the
Galactic center region. In this case the description of the pulsar
wind nebula is modified from a bow shock to a termination shock
picture (Cheng et al. 2004). However, the weak dependence of the
X-ray luminosity on the shock radius (see equ. 3) in the fast
cooling regime leads to luminosities which are not significantly
different.  Thus, pulsar wind nebulae from either termination or
bow shocks can provide a population of faint point-like sources
characterized by photon indices $\Gamma \gtrsim 1.5$ involving
non-accreting neutron stars as well as providing a potential population
of faint tail-like sources complementary to the model involving
fast moving knots associated with supernova remnants as discussed
by Bykov (2003).

Recently, TeV $\gamma$-ray emission at a luminosity higher than
$10^{35}$ ergs s$^{-1}$ from the direction of the GC
has been reported by three independent groups, Whipple
(Kosack et al. 2004), CANGAROO (Tsuchiya et al. 2004), and HESS (Aharonian et al. 2004).
The most plausible candidates suggested for this emission include the black hole Sgr $A^*$
(Aharonian \& Neronov 2005) and the compact and powerful young
supernova remnant (SNR) Sgr A East (Crocker et al. 2005). The angular scale of the
TeV source was determined by HESS to be less than a few
arc-minutes, indicating that this $\gamma$-ray source is located
in the central $\leq 10\,pc$ region (Aharonian et al. 2004). Since TeV photons are
produced by inverse Compton scattering and the efficiency is not high, typically $10^{-3}-10^{-4}$,  
it suggests that if these TeV photons are produced by PWN of MSP it requires
about $10^3$ in 10pc region even  if the inverse Compton scattering 
efficiency is as high as 10\%(Wang, et al. 2005). 
Such a high density of MSPs is very unlikely, and it is therefore, unlikely
the PWN of MSP are responsible for TeV radiation from GC.

\subsubsection{Millisecond Pulsars in Globular Clusters}

The contribution of pulsar wind nebulae to the X-ray emission of
millisecond pulsars in globular clusters is likely to differ from that
in the Galactic field.  Observationally, the recent {\em Chandra} X-ray
studies of the millisecond pulsars in 47 Tuc (Grindlay et al. 2002) reveal
that these pulsars appear to be consistent with a thermal blackbody
spectrum characterized by a temperature corresponding to an energy
of 0.2-0.3 keV and X-ray luminosities of $10^{30-31}{\rm erg\
s^{-1}}$. The emission site of this radiation is likely to be the
heated polar caps on the neutron star surface with no evidence for
emission from a pulsar wind nebula.  On the other hand, the more recent
work by Bogdanov et al. (2005) on the spectral and long-timescale variability
analyses of {\em Chandra} observations of 18 millisecond pulsars in 47
Tuc has led to the discovery that the three sources, 47 Tuc J, O and W,
exhibit a significant non-thermal component.  The photon index of
these three sources are in the range $1\pm 0.56$, $1.33\pm 0.79$, and
$1.36\pm 0.24$ respectively. Of these, only 47 W exhibits dramatic
X-ray variability as a function of orbital phase.  We note that
since 47 Tuc O lies near the center of the cluster where the number
density of X-ray sources is large its non-thermal spectrum may be
contaminated by background sources in the field.  Of the remaining
two millisecond pulsars, it is possible that the non-thermal spectral
components are produced in an intrabinary shock formed by the interaction
between the relativistic wind and matter from the stellar companion
(Bogdanov et al. 2005).  Thus, these observations  do not 
suggest a magnetospheric origin for the non thermal emission at such levels.
The much higher non-thermal X-ray luminosity from W
($\sim 2.7 \times 10^{31}$ ergs s$^{-1}$), in comparison to J ($\sim 9.3 \times
10^{30}$ ergs s$^{-1}$), may reflect the differing nature of the companion star.
A main sequence companion star nearly filling its Roche lobe and of mass
$\gtrsim 0.13 \msun$ is associated with W whereas a brown dwarf 
underfilling its Roche lobe and of mass $\lesssim 0.03 \msun$ is associated 
with J.  The X-ray luminosity of the non-thermal components from these MSPs can
be estimated following the discussion given in \S 3.3. Taking an
average spin-down power of the millisecond pulsars in 47 Tuc, $\dot E 
\sim 2\times 10^{34} {\rm erg\ s^{-1}}$, $p=2.3, \epsilon_B\sim 0.003, \epsilon_e 
\sim 0.5, \gamma_w \sim 10^6$, and $\Omega=1$, we find an non-thermal X-ray
luminosity in the band 0.1-10 keV of $4\times 10^{29} {\rm erg\ s^{-1}}$ 
for an assumed $R_s \sim 10^{11}$cm. 
If we choose $\epsilon_B\sim 0.1, \gamma_w \sim 2\times 10^5$, a larger
X-ray luminosity $\sim 4\times 10^{30} {\rm erg\ s^{-1}}$ results, 
which is closer to the observed values.
This may suggest that $\epsilon_B$ in PWN of MSPs is indeed larger.
The remaining deviation by a factor 
of 3 may suggest an underestimate of the spin down power of these 
millisecond pulsars by a factor of $\sim 2$ .

Pulsar wind nebulae are not likely to significantly 
contribute to the non-thermal
emission from millisecond pulsars in 47 Tuc since the nebula is
significantly affected by its interaction with cluster stars.
An estimate of the bow shock radius follows from equ (1).  Here,
we adopt an electron density of $\sim 0.1 {\rm cm^{-3}}$ as
inferred from the electron dispersion measures from the radio pulses
(Freire et al. 2001).  Since the millisecond pulsars lie in the
core of the cluster, we assume that the pulsar's mean velocity
is $\sim 10 {\rm km\ s^{-1}}$.  Taking the mean spin-down power of
the pulsars to be $\dot E \sim 2\times 10^{34} {\rm erg\ s^{-1}}$
(Cheng \& Taam 2003), leads to a termination radius of the pulsar
wind nebula in the case of a bow shock of $R_s \sim 2 \times 10^{18}$
cm.  This length scale is significantly larger than the average distance
between stars ($\lesssim 10^{17}$ cm) within three core radii of
47 Tuc where millisecond pulsars are primarily found in the cluster. Hence,
the interaction of stars with the nebula is likely to significantly
diminish any non-thermal non-pulsed X-ray emission associated with this
diffuse region.

This result is also expected to apply to PSR B1821-24, the brightest
millisecond pulsar in the Galaxy.  PSR B1821-24 was the first millisecond pulsar
to be discovered in a globular cluster (Lyne et al. 1987), having been found in
M28. It is characterized by a pulse period of 3.05 ms and a spin down power
of $2.2 \times 10^{36}$ ergs s$^{-1}$.  In contrast to the isolated millisecond
pulsars in 47 Tuc, its spectrum is apparently non-thermal with a power law
index $\Gamma = 1.2^{+0.15}_{-0.13}$ (see Becker et al. 2003). In addition, 
the X-ray luminosity of the diffuse component of B1821-24 is somewhat 
uncertain because of source confusion as it is located in the
core of M28.  In particular, Becker et al. (2003) obtained 
{\em Chandra}/ACIS-S spectral observations of the field of M28 and 
found 46 point sources above a limiting luminosity of $6 \times 10^{30}$ 
erg s$^{-1}$. On the other hand,
Rutledge et al. (2004), using the {\em Chandra}/HRC-S instrument to
perform absolute timing observations of B1821-24, found 7 sources
including B1821-24 within the 14".4 core radius of M28.  The time averaged
X-ray luminosity was found by Rutledge et al. (2004) to be $(4.0\pm 0.2)
\times 10^{32}$ ergs s$^{-1}$ (0.1-2.0KeV) of which 15\% (or $6
\times 10^{31}$ ergs s$^{-1}$) was attributed to the nonpulsed component.
Although the spectrum of this component is not known, its luminosity
level corresponds to a very small fraction ($\sim 3 \times 10^{-5}$)
of the spin down power of B1821-24 and could originate from  
source confusion or from some residual nearly isotropic magnetospheric
X-ray emission.

Based on the interpretation for the lack of non-thermal emission for
the isolated millisecond pulsars in 47 Tuc, it is likely that stellar
interactions with the relativistic wind from B1821-24 would significantly
affect the establishment of a PWN.
Specifically, the termination shock radius
is estimated to be of the order of 1 pc, while the high core
density in M28 ($\sim 10^5$ stars pc$^{-3}$) indicates that the average
distance between stars is 50 times smaller. Hence, the results 
described above suggest that the low level of non-thermal non-pulsed 
X-ray emission from PSR B1821-24 may represent the contribution from 
the pulsar's magnetosphere.

\section{DISCUSSION}

For the 8 millisecond pulsars in the Galactic field for which the 
2-10 keV total X-ray luminosity can be determined, 2 have been published 
with the pulsed and non-pulsed emission components. 
In this paper, we have interpreted the non-pulsed component as arising
from a PWN. Evidence in support of such an
interpretation is provided by the observation of diffuse 
X-ray emission in the millisecond pulsar, B1957+20 (Stappers et al.
2003). For the majority of millisecond pulsars where X-ray
emission has been detected (see Table 1), the X-ray emission is
not spatially resolved. Since the X-ray luminosity associated
with the pulsed emission from the pulsar's magnetosphere is
expected to be a relatively small fraction of the total X-ray
luminosity, $L^{\rm pul}_X/ \dot E \sim 3\times 10^{-4}$ in the
model of Cheng et al. (1998) and Zhang \& Cheng (2003), it is
likely that the wind nebulae significantly contribute to the
observed X-ray radiation. 

A simple one-zone model for the X-ray emission from the nebula 
(Cheng, Taam \& Wang 2004),
based on the work of Chevalier (2000), indicates that the
efficiency for the conversion of rotational energy of the neutron
star to X-ray emission can be significantly higher in the post
shocked region of the wind nebula than in the neutron star
magnetosphere. In this model, the X-ray luminosities primarily
depend upon the spin down power of the pulsar and particle and
magnetic field energy densities behind the shock and only weakly
dependent on the pulsar's velocity and the number density of the
surrounding interstellar medium.  The X-ray
emission is characteristically non-thermal and described by
a power law with photon index lying in the range between 1.5 and
2.5, with the efficiency of conversion of spin down power to X-ray
luminosity greater in the fast cooling regime
(where $\Gamma \sim 2-2.5$) than in the slow cooling regime (where
$\Gamma \sim 1.5-2$).  In our interpretation, the X-ray source may
be either point-like or tail-like in appearance, and we interpret
the elongated tail observed in several faint X-ray sources as the
distance traversed by the pulsar within the synchrotron emission
cooling timescale.  Pulsars moving more rapidly than 100 km
s$^{-1}$ in regions of the interstellar medium characterized by
magnetic field strengths $\lesssim 0.1$ mG are likely to exhibit a
tail-like structure even at distances comparable to the Galactic
center. For example, for $B \sim 0.03$ mG and $v_p \sim 150$ km
s$^{-1}$ the tail can extend to $\sim 3 \times 10^{17}$ cm, a
length scale comparable to that of elongated sources observed in
the deep {\em Chandra} surveys of the Galactic center (see Wang et
al. 2002b, Muno et al. 2003).  The majority of the sources, on the
other hand, are likely to have much smaller sized nebulae, making
them appear point-like. For typical spin parameters of millisecond
pulsars, the X-ray luminosity level can be as high as $\sim
10^{31} - 10^{32}$ ergs s$^{-1}$, perhaps making them an important
contributor to the  faint X-ray source population associated with 
neutron stars in deep X-ray surveys.

Because the emission from PWN is spatially
extended and can exceed the luminosity of the neutron star, the
measured emission from a pulsar may be a function of the size of
the region studied.  As the size depends on the magnetic field and
number density of the pulsar's environment, we illustrate their
range by listing in Table 2 the estimated size of the pulsar wind
nebulae surrounding millisecond pulsars, for physical conditions
in the Galactic center and in B1957+20.  For comparison, the highest 
spatial resolution afforded by the {\em Chandra} observations is also 
listed showing that the PWN can be entirely contained 
within or  extend beyond the aperture radius respectively. Although the 
simple one-zone model is inadequate to predict the spatial variation 
of the photon power law index and luminosity, it is expected that 
the X-ray properties of pulsars surrounded by wind nebulae are likely 
to be a function of the observed spatial resolution. 

We have suggested that the contribution of the PWN to the
non-pulsed non thermal emission of millisecond pulsars in the cores of
globular clusters is likely reduced due to the effect of interaction of 
stars with it. 
Hence, the non-thermal X-ray contribution due to the PWN 
to the total X-ray emission may not be important in comparison to millisecond 
pulsars in the Galactic field. For those millisecond pulsars with 
a binary companion in globular cluster systems, the emission from an 
intrabinary shock can be an important contributor to the non-thermal 
non-pulsed X-ray emission. In particular, it has been found (see equation  
7) that the non-thermal X-ray emission from an intrabinary shock in 
a binary millisecond pulsar system is very sensitive to the spin down 
power. For example, a spin down power of $10^{34}$ ergs s$^{-1}$ would 
yield a non-thermal X-ray luminosity of $\sim 3 \times 10^{29}$ erg 
s$^{-1}$ for $p$ = 2.2, making them difficult to detect at the present
time. Thus, the lack of non-thermal X-ray emission from the 
majority of binary millisecond pulsars in 47 Tuc suggests that 
$\dot E \lesssim 10^{34}$ ergs s$^{-1}$ for the majority of these 
binary millisecond pulsars. 

In contrast, the non-thermal emission from isolated millisecond pulsars in 
globular clusters may have a contribution from the magnetosphere.  The 
identification of a non-thermal non-pulsed component in globular cluster 
millisecond pulsars uncontaminated by source confusion could be
especially important in constraining the flux level of such a component
in pulsar emission models.  Some unpulsed X-ray emission 
is expected if the distribution of the pitch angles of charged 
particles in the magnetosphere is uniform.  In this case, the synchrotron 
radiation from  particles with very large pitch angle, say larger than 80$^{\circ}$, 
will tend to radiate isotropically. The luminosity of this component can be 
estimated as $L_x(unpulsed)/\dot{E}= (L_x(unpulsed)/L_x(pulsed))
(L_x(pulsed)/\dot{E})\sim cos(80^{\circ}) 3\times 10^{-4} \sim 
5\times 10^{-5}$. Here we have assumed that $(L_x(unpulsed)/L_x(pulsed))$ is
given by the ratio of their angular distribution $\sim cos(80^{\circ})$
and $(L_x(pulsed)/\dot{E})\sim 3\times 10^{-4}$ is estimated by Cheng \& Zhang (1999).
Such a level of emission is comparable to the 
off-pulse X-ray emission detected from PSR J0218+4232 and PSR B1821-24 (e.g. 
Mineo et al. 2000).  
We remark that other emission sites for non-thermal non-pulsed X-rays
have been suggested. For example, Cheung \& Cheng (1994) have argued that
pair production can still continue beyond the light cylinder, where the magnetic 
field lines are twisted and the pitch angles of the created pairs are large.
Consequently the radiation of relativistic electron and positron pairs
will be not beamed. The characteristic photon
index is $\sim 1.5$. Without high angular
resolution, it will be difficult to distinguish the origin of the non-pulsed
non-thermal X-ray emission. In particular, the photon index of the X-ray
emission from the shock region in the slow cooling regime cannot be differentiated
from the unpulsed emission originating in the magnetosphere, making
direct observational determination of the non-thermal X-ray emission location
challenging.

Our model X-ray luminosity depends on several shock parameters. In particular, 
$\epsilon_B$ and $\gamma_w$ are poorly determined. However, both equation 3 
and equation 5 do not sensitively depend on $\gamma_w$ if $p$ is around 2. 
Again the dependence on $\epsilon_B$ in equation 3 is also very weak. However, this parameter
become important for the case of an intrabinary shock. In fact, by comparing with systems with 
a contribution from an intrabinary shock the inferred $\epsilon_B$ is significant larger
than that of the Crab nebula. For example, $\epsilon_B \sim$ 0.02 can match the observed
data of PSR1957+20. Although this is significantly higher than inferred for the 
Crab nebula, it is consistent with the value obtained in 
PSR 1259-63 (Tavani \& Arons 1997).  These results taken together point 
to the importance of the environmental properties of the surrounding medium on estimates 
of $\epsilon_B$.

In this paper, we have further assumed that $B_d\cdot \Omega<$0 in estimating
the potential drop, where $B_d$ and $\Omega$ are the dipole magnetic field and 
the angular velocity of the star respectively. In the case of $B_d\cdot \Omega>$0, 
Harding, Usov \& Muslimov (2005) showed that the potential drop in the polar gap 
could be much larger than that estimated in \S 3.1, however, the X-ray emission 
associated with the polar gap emission will be pulsed and also correlated with 
the gamma-ray emission. Therefore this may provide a test to discriminate these two
cases.

Finally, the pulsar wind interacting with a companion star 
and/or the interstellar medium may also contribute to the hard X-ray
emission associated with soft X-ray transients in quiesence if a
millisecond pulsar is activated during temporary states when the
accretion of mass is significantly reduced or prevented by a
rapidly rotating magnetosphere (Illarionov \& Sunyaev 1975). 
We suggest that observations be carried out to search for the
non-thermal radio emission that would accompany the hard X-ray
emission associated with the synchrotron radiation process to
confirm the viability of such an interpretation.  If detected,
millisecond pulsars in isolation and in binary systems may be more
important contributors to the faint X-ray source population in the
Galaxy than has hitherto been considered.  Such an association may
prove useful as a means of identifying candidate objects for
further radio pulse timing searches of new millisecond pulsars.

\acknowledgements We are grateful for useful comments from the 
referee and discussions with Drs. Craig Heinke and Werner Becker.
This work is partially supported by the NSF through grant 
AST-0200876, by a RGC grant of the Hong Kong Government, by 
the National Natural Science Foundation of China under grant 
10273011 and by the Theoretical Institute of Advanced Research in 
Astrophysics (TIARA) in Taiwan.

\begin{figure}

\plotone{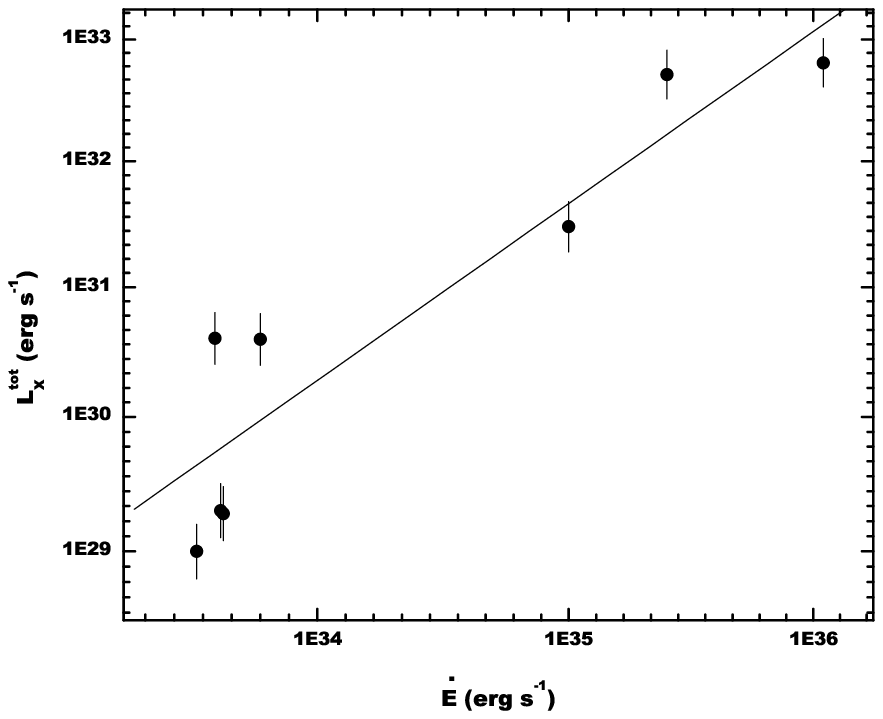} 
\caption{The total X-ray luminosity (2-10 keV)
versus spin-down power of 8 millisecond pulsars. The solid line
is the best fitting with a function with $L_{\rm x} \propto 
\dot E^{1.39\pm 0.08}$ ergs s$^{-1}$.}
\end{figure}

\begin{table}

\caption{X-ray properties of some millisecond pulsars }
\begin{center}
\scriptsize{
\begin{tabular}{l c c c c c c c c l}
\tableline \tableline
PSR & $P$(ms) & $\dot P({\rm s\ s^{-1}})$ &
$d$ & $\dot E$ & $L_{\rm X,tot}$ & $L_{\rm X,pul}$ &
$L_{\rm X,npul}$ & $\Gamma$ & Reference \\
\tableline B1937+21 & 1.56 & 1.05$\times 10^{-19}$ & 3.6  &
1.1$\times 10^{36}$ & 6.4$\times 10^{32}$& 4.6$\times 10^{32}$
& 1.8$\times 10^{31}$ & ${1.21\pm 0.15^a\over 3.3\pm 0.5^b}$ & 1 \\
J2124-3358 & 4.93 &  1.3$\times 10^{-20}$ & 0.25 & 4.3$\times
10^{33}$ & 1.9$\times 10^{29}$ & & & $2.8\pm 0.6^d$ & 2 \\
J0437-47  & 5.8&  1.9$\times10^{-20}$  & 0.139$^c$ & 4.2$\times
10^{33}$& 2.0$\times 10^{29}$ & & & $2.2\pm 0.3$ & 3\\
J0030+0451 & 4.87& 1.0$\times 10^{-20}$  & 0.23 &
3.4$\times 10^{33}$ & 1.0$\times 10^{29}$ &   & & $3\pm 0.4$ &4 \\
J0218+4232 & 2.3 & 7.5$\times 10^{-20}$ & 2.7$^c$ & 2.5$\times10^{35}$
& $5.12 \times 10^{32}$ & $4.6 \times 10^{32}$ & $3 \times 10^{31}$ &
${0.61\pm 0.32^a \over 1.17\pm 0.37^b}$
& 5, 6 \\
J0751+1807 & 3.48& 8$\times 10^{-21}$ & $1.1^c$ &  6$\times 10^{33}$ &
3.93$\times 10^{30}$ &  & &  $1.59 \pm 0.2$   & 7\\
J1012+5307 & 5.3 &  1.5$\times 10^{-20}$ & 0.77$^c$ &
4$\times 10^{33}$ & 4.0$\times 10^{30}$ & & &  $1.78 \pm 0.36$   & 7 \\
B1957+20  &  1.6 & 1$\times10^{-20}$ & $2.5^c$  & 1$\times 10^{35}$ &
3.0$\times 10^{31}$ & & & 1.9$\pm 0.5$ & 7 \\
\tableline

\end{tabular}}

\end{center}

\tablecomments{The first column PSR is the pulsar name, $P$ is the
spin period, $\dot P$ is the period derivative, $d$ is the
distance of the pulsar in units of kpc. The luminosity is in units
of erg\ s$^{-1}$. $\dot E$ is the pulsar's spin-down power.
$L_{\rm X,tot}$ is the total X-ray luminosity, $L_{\rm X,pul}$ is
just the pulsed X-ray luminosity, $L_{\rm X,npul}$ is the non
pulsed luminosity. 
In calculating the pulsed X-ray luminosity, the solid angle 
is assumed to be unity. $\Gamma$ is the photon index. The luminosities
are taken from the observations of {\em ASCA, BeppoSAX,
XMM-Newton} and {\em Chandra} in the energy range of 2-10 keV. 
We have approximated errors in log($L_x$) as $\pm$ 0.2.
\\
$^a$ $\Gamma$ of the pulsed components ; \\
$^b$ $\Gamma$ of the nonpulsed components; \\
$^c$ Accurate distance estimates (Bogdanov et al. 2005). Other
estimates are obtained mainly from dispersion measure together
with the Cordes \& Lazio (2002) electron density model and are
rather uncertain. \\
$^d$ $\Gamma=2.2\pm 0.3$ given by Hui \& Becker (2005)
Reference: 1. Nicastro et al. 2004; 2.
Sakurai et al. 2001; 3. Zavlin et al. 2002; 4. Becker \&
Aschenbach 2002;  5. Mineo et al. 2000; 6. Webb et al. 2004a; 7. Webb et al. 2004b; 8. Stappers et al. 2003}
\end{table}

\begin{table}

\caption{Relative size of pulsar wind nebulae with respect to
Chandra resolution for millisecond pulsars in two different environments}
\begin{center}
\begin{tabular}{l c c c c c c c l}
\tableline \tableline \ & $D$ &  $n$ & $B$ & $v_p^\dag$ &
$\delta\theta$ & $R_{\rm obs}$ & $R_s$  \\
\ & (kpc) & (${\rm cm^{-3}}$) & G & $({\rm km\ s^{-1}})$ & \ & (cm) &
 (cm) &   \\
\tableline GC & 8.5 & $10^2$  & $10^{-4}$ & 130 & 1" & $6\times
10^{16}$ & $2\times 10^{15}$  \\
1957+20 & 1.5 & 1 & $10^{-5}$ & 220 & 1" & $1\times 10^{16}$  &
$4\times 10^{16}$  \\
\tableline

\end{tabular}

\end{center}

\tablecomments{Millisecond pulsars in the Galactic Center and 
PSR B1957+20.  $D$ is the distance, $n$ is the number density
in the medium surrounding the pulsar, $B$ is the magnetic field in
the interstellar medium, $v_p$ is the pulsar proper motion
velocity, $\delta\theta$ is the detection angular limit in the
different observations, $R_{\rm obs}\sim D\delta\theta/2$ is the
radius of the aperture, and $R_s$ is the predicted shock radius.\\
$^\dag$ For the Galactic center, we take a spin down power of $2 
\times 10^{34}$ ergs s$^{-1}$ and an average
velocity of pulsars of 130 km s$^{-1}$, which is lower than the escape speed
from this region.}
\end{table}

\begin{references}
\reference{} Aharonian, F.A., et al. 2004, \aap, 425, L13
\reference{} Aharonian, F.A., \& Neronov, A., 2005, \apj, 619, 306
\reference{} Alpar M.A., Cheng A.F., Ruderman M.A., Shaham J., 1982, Nat, 300, 728
\reference{} Arons, J. 1981, \apj, 284, 1099
\reference{} Arons, J. \& Tavani, M., 1993, \apj, 403, 249
\reference{} Arzoumanian, A., Fruchter, A. S., \& Taylor, J. H.,
1994, \apj, 426, L85
\reference{} Backer, D. C., Kulkarni, S. R., Heiles, C., Davis, M. M., \&
Goss, W. M. 1982, \nat, 300, 615
\reference{} Bailes, M. et al. 1997, \apj, 481, 386
\reference{} Beck, R., Shukurov, A., Sokoloff, D., Wielebinski, R. 2003, \aap, 
411, 99
\reference{} Becker, W. \& Aschenbach, B. 2002, ``Neutron
Stars, Pulsars and Supernova Remnants'' eds. W. Becker, H. Lesch
\& J. Tr\"umper, MPE Report 278, 64
\reference{} Becker, W. \& Tr\"umper, J. 1993, \nat, 365, 528
\reference{} Becker, W. \& Tr\"umper, J. 1997, \aap, 326, 682
\reference{} Becker, W. \& Tr\"umper, J. 1998, IAU circular 6829
\reference{} Becker, W. \& Tr\"umper, J. 1999, \aap, 341, 803
\reference{} Becker, W. et al. 2003, \apj, 594, 798
\reference{} Becker, W. et al. 2005, \apj, submitted
\reference{} Bednarz, J. \& Ostrowski, M. 1998, Phys. Rev. Lett.,
80, 3911
\reference{} Belczynski, K, \& Taam, R. E. 2004, \apj, 616, 1159
\reference{} Blondin, J.M., Chevalier, R.A.  \& Frierson, D.M. 2001, \apj, 563, 806
\reference{} Bogdanov, S., Grindlay, J. E., \& van den Berg M. 2005, 
\apj, 630, 1029
\reference{} Bogdanov, S., Grindlay, J. E., Heinke, C., Camilo, F.
Freire, P. C. , Becker, W. 2005, submitted to ApJ
\reference{} Bykov, A. M. 2003, \aap, 410, L5
\reference{} Bucciantini, N. 2002, \aap, 387, 1066
\reference{} Campana, S., Colpi, M., Mereghetti, S., Stella, L.,
\& Tavani, M. 1998, \aapr, 8. 279
\reference{} Campana, S. et al. 2004, \apj, 614, L49
\reference{} Campana, S., \& Stella, L. 2003, astro-ph/0309811
\reference{} Cheng, A. F., \& Ruderman, M. A. 1980, \apj, 235, 576
\reference{} Cheng, K. S., Gil, J. \& Zhang, L. 1998, \apj, 493,
L35
\reference{} Cheng, K. S., \& Taam, R.E. 2003, \apj, 598, 1207
\reference{} Cheng, K. S., Taam, R. E., and Wang, W. 2004, \apj,
617, 480
\reference{} Cheng, K. S. \& Zhang, L. 1999, \apj, 515, 337
\reference{} Cheung, W.M., \& Cheng, K. S. 1994, \apjs, 90, 827
\reference{} Chevalier, R. A. 2000, \apj, 539, L45
\reference{} Cordes, J.M. \& Lazio, T.J.W. 2002, astro-ph/0207156
\reference{} Crocker, R.M., et al., 2005, \apj, 622, 892
\reference{} de Jager, O.C. \& Harding, A.K. 1992, \apj, 396, 161
\reference{} Freire, P. C. et al. 2001, \apj, 557, L105
\reference{} Fruchter, A. S., Stinebring, D. R., \& Taylor, J. H.
1988, \nat, 333, 23
\reference{} Gaensler, B.M. 2004, in IAU Symposium 218 "Young Neutron Stars and
Their Environments" eds F. Camilo \& B.M. Gaensler (astro-ph/0405290)
\reference{} Goldreich, P.,  \& Julian, W.H. 1969, \apj, 157, 859
\reference{} Grindlay, J. E. et al. 2002, \apj, 581, 470
\reference{} Harding, A. K., \& Muslimov, A.G. 2001, \apj, 556, 987
\reference{} Harding, A. K., \& Muslimov, A.G. 2002, \apj, 568, 862
\reference{} Harding, A. K., Usov, V.V., \& Muslimov, A.G. 2005, \apj, 622, 531
\reference{} Helfand, D. J., Gotthelf, E. V., \& Halpern, J. P.
2001, \apj, 556, 380
\reference{} Hui, C.Y. \& Becker, W. 2005, \aap, submitted for publication
\reference{} Illarionov, A., \& Sunyaev, R. 1975, \aap, 39, 185
\reference{} in 't Zand, J. J. M. et al. 2001, \aap, 372, 916
\reference{} Jonker, P. G., Wijnands, R., \& van der Klis 2004,
\mnras, 349, 94
\reference{} Jones, P. 1980, \mnras, 192, 847
\reference{} Kennel, C. F. \& Coroniti, F. V. 1984, \apj, 283, 694
\reference{} Komissarov, S.S. \& Lyubarksi, Y.E. 2003, \mnras, 344, L93
\reference{} Kosack, K. et al., 2004, \apj, 608, L97
\reference{} Kulkarni, S. R., \& Hester, J. J. 1988, \nat, 335,
801
\reference{} Lemoine, M. \& Pelletier, G. 2003, \apj, 589, L73
\reference{} Lu, F. Wang, Q.D. \& Lang, C. 2003, \aj, 126, 319
\reference{} Lyne, A. G., Brinklow, A., Middleditch, J., Kulkarni, S. R., 
Backer, D. C., \& Clifton, T. R. 1987, \nat, 328, 399
\reference{} Lyne, A. G., Manchester, R. N., Lorimer, D. R., Bailes, M.,
D'Amico, N., Tauris,
T. M., Johnston, S., Bell, J. F., \& Nicastro, L. 1998, \mnras, 295, 743
\reference{} Mineo, T. et al. 2000, A\&A, 355, 1053
\reference{} Morris, M. 1994, The Nuclei of Galaxies, eds.
Reinhard Genzel \& Andres I. Harris (Kluwer, Dordrecht), p. 185
\reference{} Muno, M. P., et al. 2003, \apj, 589, 225
\reference{} Muno, M. P., et al. 2004, \apj, 613, 1179
\reference{} Nicastro, L. et al. 2004, \aap, 413, 1065
\reference{} $\ddot {\rm O}$gelman, H. 1995, in The Lives of Neutron Stars, eds.
A. Alpar, U. Kilizoglu \& J. van Paradijs, (Kluwer, Dordrecht), p. 101
\reference{} Park, S., Muno, M. P., Baganoff, F. K., Maeda, Y.,
Morris, M., Howard, C., Bautz, M. W., \& Garmire, G. P. 2004,
\apj, 603, 548
\reference{} Peterson, L. E. \& Jacobson, A. S. 1970, PASP, 82,
412
\reference{} Phinney, E. S., \& Kulkarni, S. R. 1994, \araa, 32, 591
\reference{} Psaltis, D.,  \& Chakrabarty, D. 1999, \apj, 521, 332
\reference{} Ruderman, M., Shaham, J. \& Tavani, M., 1989, \apj, 336, 507
\reference{} Rutledge, R. E., Fox, D. W., Kulkarni, S. R., Jacoby, B. A., Cognard, 
I., Backer, D. C., \& Murray, S. S. 2004, \apj, 613, 522
\reference{} Sakurai, I.  et al. 2001, \pasj, 53, 535
\reference{} Saito, Y. et al.1997b, \apj, 477, L37
\reference{} Saito, Y. 1998, Ph.D.  Thesis, Univ. of Tokyo
\reference{} Sefako, R.R. \& DeJager, O.C. 2003, \apj, 593, 1013
\reference{} Seward, F. D, \& Wang, Z. R. 1988, \apj, 332, 199
\reference{} Stappers, B. W., Gaensler, B. M., Kaspi, V. M., van
der Klis, M., \& Lewin, W. H. G. 2003, Science, 299, 1372
\reference{} Stella, L. et al. 1994, ApJ, 423, L47
\reference{} Tavani, M., Arons, J. \& Kapsi, V. M. 1994, 433, L37
\reference{} Tavani, M. \& Arons, J. 1997, 477, 439
\reference{} Toscano, M. et al. 1999, \mnras, 307, 925
\reference{} Tsuchiya, K. et al. 2004, \apj, 606, L115
\reference{} Tsuruta, S. 1998, Phy. Rep., 292 1
\reference{} van der Swaluw, E. \aap, 404, 939
\reference{} Wang, Q.D., Gotthelf, E.V., \& Lang, C.C. 2002, \nat, 415, 148
\reference{} Wang, Q.D., Lu, F. \& Lang, C. 2002, \apj, 581, 1148
\reference{} Wang, W., Jiang, Z.J., Pun, J.C.S. \& Cheng, K.S. 2005, \mnras, 360, 646
\reference{} Webb, N. A., Olive, J.-F \& Barret, D. 2004a, \aap, 417, 181
\reference{} Webb, N. A., Olive, J.-F, Barret, D., Kramer, M., Cognard, I. \& L$\ddot{o}$hmer, O., 2004b, \aap, 419, 
269
\reference{} Weisskopf M. C. et al. 2000, \apj, 536, L81
\reference{} Wijnands, R., \& van der Klis, M. 1998, \nat, 394, 344
\reference{} Zavlin, V.E. et al., 2002, \apj, 569, 894
\reference{} Zhang, B., \& Harding, A. K. 2000, \apj, 532, 1150
\reference{} Zhang, L. \& Cheng, K.S. 2003, \aap, 398, 639
\end{references}
\end{document}